\documentclass[sigconf]{acmart}

\acmConference[ICSE 2024]{46th International Conference on Software Engineering}{April}{Lisbon, Portugal}

\usepackage{latexsym}

\usepackage[T1]{fontenc}

\usepackage[utf8]{inputenc}

\usepackage{microtype}

\usepackage{graphicx}
\usepackage{tabularx}
\usepackage{booktabs}
\usepackage{amsmath}
\usepackage{algorithm}
\usepackage[noend]{algorithmic}
\usepackage{xcolor}
\usepackage{soul}
\usepackage{enumitem}
\usepackage{amsthm} %
\usepackage{mdframed}

\usepackage{enumitem} %
\usepackage{courier} %
\usepackage{lipsum} %
\usepackage{natbib}
\usepackage{xurl}
\definecolor{myred}{RGB}{242,172,185}
\definecolor{mygreen}{RGB}{173,223,179}

\newcommand{\hlred}[1]{%
    {%
    \sethlcolor{myred}%
    \hl{#1}%
    }%
}
\newcommand{\hlgreen}[1]{%
    {%
    \sethlcolor{mygreen}%
    \hl{#1}%
    }%
}

\usepackage{newfloat}
\usepackage{subcaption}
\usepackage{listings}
\lstdefinestyle{prompt}
{
    xrightmargin=1em,
    xleftmargin=1em,
    basicstyle=\small\ttfamily,
    keywordstyle=\color{blue},
    morekeywords={@output,@input,string,bool,string[],int,Tuple,<,>,language,@values,using,Regex},
}
\usepackage[frozencache,cachedir=minted-cache]{minted} 
\usepackage[para,online,flushleft]{threeparttable}

\usepackage{tikz}
\newcommand*\circled[1]{\textcircled{\raisebox{-0.8pt}{#1}}}

\definecolor{ahcolor}{rgb}{0.1,0.7,0.8}
\definecolor{vlcolor}{rgb}{0.9,0.1,0.1}
\definecolor{gcolor}{rgb}{0.7,0.3,0.7}
\definecolor{cpcolor}{rgb}{0.3,0.3,0.7}
\definecolor{mscolor}{RGB}{8, 102, 3}
\definecolor{ascolor}{RGB}{163,96,50}
\definecolor{akcolor}{rgb}{0.36, 0.54, 0.66}

\title{Semantically Aligned Question and Code Generation for Automated Insight Generation}

\author{Ananya Singha } 
\email{t-asingha@microsoft.com}
\affiliation{%
  \institution{Microsoft}
  \country{India}}
  
\author{Bhavya Chopra}
\email{t-bhchopra@microsoft.com}
\affiliation{%
  \institution{Microsoft}
  \country{India}}

\author{Anirudh Khatry}
\email{t-anikhatry@microsoft.com}
\affiliation{%
  \institution{Microsoft}
  \country{India}}
  
\author{Sumit Gulwani}  
\email{sumitg@microsoft.com}
\affiliation{%
  \institution{Microsoft}
  \country{USA}}
  
\author{Austin Z. Henley}
\email{azh321@gmail.com}
\affiliation{%
  \institution{Microsoft}
  \country{USA}}
  
\author{Vu Le} 
\email{levu@microsoft.com}
\affiliation{%
\institution{Microsoft}
\country{USA}}
  
 \author{Chris Parnin}  
 \email{chrisparnin@microsoft.com} 
 \affiliation{%
\institution{Microsoft}
\country{USA}}

  \author{Mukul Singh}
  \email{singhmukul@microsoft.com}
 \affiliation{%
  \institution{Microsoft}
  \country{India}}

 \author{Gust Verbruggen} 
  \email{gverbruggen@microsoft.com}
 \affiliation{%
\institution{Microsoft}
\country{Belgium}}

\renewcommand{\shortauthors}{Ananya Singha et al.}

\copyrightyear{2024}
\acmYear{2024}
\setcopyright{acmlicensed}\acmConference[LLM4Code '24]{2024 International Workshop on Large Language Models for Code}{April 20, 2024}{Lisbon, Portugal}
\acmBooktitle{2024 International Workshop on Large Language Models for Code (LLM4Code '24), April 20, 2024, Lisbon, Portugal}
\acmDOI{10.1145/3643795.3648381}
\acmISBN{979-8-4007-0579-3/24/04}

\begin{document}

\begin{abstract}
Automated insight generation is a common tactic for helping knowledge workers, such as data scientists, to quickly understand the potential value of new and unfamiliar data. Unfortunately, automated insights produced by large-language models can generate code that does not correctly correspond (or \emph{align}) to the insight. In this paper, we leverage the semantic knowledge of large language models to generate targeted and insightful questions about data and the corresponding code to answer those questions. Then through an empirical study on data from Open-WikiTable, we show that embeddings can be effectively used for filtering out semantically unaligned pairs of question and code. Additionally, we found that generating questions and code together yields more diverse questions.

\end{abstract}
\maketitle
\section{Introduction}

In the \emph{Age of Copilots}\footnote{\url{https://www.forbes.com/sites/forbestechcouncil/2023/09/08/the-age-of-co-pilots}}, numerous companies have developed \emph{copilots} that leverage large-language models (LLMs) to answer and perform tasks directly in products.
To facilitate user interaction with the product copilot, a common tactic is to automatically generate suggested questions or followups.
This tactic is also common in exploratory data analysis, where automated insights are generated to help data scientists initiate interaction with data~\cite{ma2021metainsight}.
More generally, a \emph{good question} can be an effective tool for human knowledge acquisition~\cite{graesser2004autotutor}, and effective problem-solving~\cite{shridhar-etal-2022-automatic}.

\begin{figure}[ht]
\centering
\scriptsize
\begin{tabular}{@{}lrllc@{}}
\toprule
Outcome  & Year & Championship               & Opponent in final & Score \\ 
\midrule

Winner    & 1990   & World Snooker Ch. & Jimmy White              & 18--12  \\
Winner    & 1990   & Grand Prix (2)             & Nigel Bond               & 10--5   \\
Winner    & 1987   & Grand Prix                 & Dennis Taylor            & 10--7   \\
Winner    & 1990   & Asian Open (2)             & Dennis Taylor            & 9--3    \\
Winner    & 1990   & Dubai Classic (2)          & Steve Davis              & 9--1    \\
Winner    & 1990   & UK Championship (2)        & Steve Davis              & 16--15  \\
Winner    & 1988   & British Open               & Mike Hallett             & 13--2   \\
Runner-up & 1988   & UK Championship            & Doug Mountjoy            & 12--16  \\
Winner    & 1989   & Asian Open                 & James Wattana            & 9--2    \\
Winner    & 1989   & Dubai Classic              & Doug Mountjoy            & 9--2    \\
Winner    & 1989   & UK Championship            & Steve Davis              & 16--12  \\

Runner-up & 1990   & European Open              & John Parrott             & 6--10   \\

Runner-up & 1991   & Classic                    & Jimmy White              & 4--10 \\

Runner-up & 1989   & International Open         & Steve Davis              & 4--9    \\
\bottomrule
\end{tabular}

    \caption{A table displaying the snooker championship results from the Open-WikiTable corpus.}

\label{tab:motiv_scenario}
\end{figure}

In this paper, we leverage large language models \cite{gpt3} to generate \textbf{semantically-aligned} question and code pairs for supporting automated insight generation.
Data scientists (and other users) often explore a new dataset, by using tools, writing queries, or code. 
Unfortunately this process takes up a significant part of data-driven workflows~\cite{Dasu2002MiningDS, guo2013data}---understanding how to select and combine operations such as  sorting, filtering, reshaping, grouping, joining, and pivoting data can be time-consuming.
This problem can be made even worse if a data scientist is mislead by using incorrect code generated by an LLM.

For example, consider a table with snooker data for a specific player, as shown in Figure~\ref{tab:motiv_scenario}.
Following are a few examples of the automated insights generated by a LLM.

\noindent{\hlgreen{ \textbf{In which championship did the player face the toughest competition based on the average frames played in the finals?}}}
\begin{minted}[fontsize=\small]{Python}
> (df.groupby('Championship')['Score']
     .apply(lambda x: 
        sum(int(score.split('-')[0]) + 
        int(score.split('-')[1]) 
        for score in x) / len(x))
     .idxmax()
'UK Championship (2)'  
\end{minted}
\smallskip
\noindent{\hlgreen{\textbf{How many championships did the player win consecutively?}}}
\begin{minted}[fontsize=\small,breaklines]{Python}
> df.sort_values(by=["Year"])
    .groupby(
        (df['Outcome'] != df['Outcome'].shift())
        .cumsum())['Outcome']
    .transform('size')
    .max()
5 
\end{minted}

The above insights, highlighted in green, are useful insights that help in understanding different segments of the tabular data and are beneficial to provide a head start with code that can be used and extended. Unfortunately, the same approach, without any additional filtering, also generates misleading and incorrect insights or code, potentially negating the value of the entire approach: 

\noindent{\hlred{\textbf{What is the overall trend in the player's performance over the years? }}}
\begin{minted}[fontsize=\small,breaklines]{Python}
>>> 'Increasing' if 
    df['Outcome'].eq('Winner').sum() > 
    df['Outcome'].eq('Runner-up').sum() else 'Decreasing' if 
    df['Outcome'].eq('Winner').sum() < 
    df['Outcome'].eq('Runner-up').sum() 
    else 'Fluctuating'
'Increasing'
\end{minted}
\begin{itemize}
    \item[$\rightarrow$]While the insight is interesting, the code does not appropriately account for multiple championships in years.
\end{itemize}
\smallskip
\noindent{\hlred{\textbf{What is the average score in the finals when the player wins compared to when they are the runner-up?}}}
\begin{minted}[fontsize=\small,breaklines]{Python}
>>> df[df['Outcome'] == 'Winner']['Score']
    .apply(lambda x: 
        sum(map(int, x.split('--'))))
    .mean()
18.0
\end{minted}
\begin{itemize}
    \item[$\rightarrow$]The insight is computing interesting stats about the data but the calculation is incorrect. The calculation must take the first part of the score after splitting rather than both.
\end{itemize}

We conducted an empirical study to understand how we can correctly identify semantically-aligned question and code amongst all generated pairs.
First, we conduct a user study to understand the relevance of the question and code pairs we generate using LLMs.
Then, to determine if questions and code align, we compare multiple models on their ability to rate the correctness of (question, code) pairs.
Finally, to measure diversity, we compute the edit distance between generated code snippets after masking constants.
Based on our study we found that: (1) users found most generated code pairs to be interesting and meaningful for their work, (2) a semantic alignment classifier based on code embeddings performs on par with GPT-4 \cite{gpt4} on a human-annotated dataset and (3) generating questions and code together gives more diverse inspirations.

\section{Related Work}

LLM are incrementally used for tasks that involve generating executable code given some textual context. 
This marks one of the challenging problems because, unlike text, sequentially generated code has to maintain its semantic and syntactical correctness. 
Several fine-tuning and training efforts have been carried out over the years to enhance the natural language to code generation capabilities. 
For example, CodeBERT (BERT), CodeT5 (T5) and Codex (GPT-3) have been fine-tuned for text-to-code tasks for multiple domains like pandas, Java etc. 
Huge training datsets, bigger models, expert tuning approaches, enriched training objectives specializing in code context have led to improvements from the baselines in terms of lower syntactical errors. 
Though semantics still remains a concern \cite{zan2023large} 
and there have been fewer efforts in generating similar code and natural language together for tabular data analysis.

An integral part of the data science workflow involved analysing the tabular data. 
Data analysis tasks such as data cleaning~\cite{huang2018auto}, formatting~\cite{singh2022cornet}, transformation~\cite{yan2020auto}, and visualization~\cite{lux} have been explored and collectively fall under the workflow of Exploratory Data Analysis (EDA).
Previously these tasks where carried out by a symbolic setup like MetaInsights \cite{ma2021metainsight} which mine common data pattern to show insights. Similarly, LLM are also used as an orchestrator along with the symbolic systems to rank and summaries insights~\cite{ma2023demonstration}. Though none of these works focus on the alignment issue between the question and the generate code.

Finally, our work is closely related to the research on automatically generating question and answers in various domains, and additional challenges associated with alignment~\cite{cambronero2019deep, wang2021codet5}, 
and diversity~\cite{yang-etal-2021-diversity-consistency} of results.
In contrast to these approaches, we specialize our approach in the table code generation domain and to handle tabular/code specific issues, leverage LLMs to derive semantic table and column information during synthesis, and leverage embeddings to train cost-effective classifiers for measuring alignment.

\section{Proposed Method}

\begin{figure*}[htb]
\centering
\includegraphics[width=1\textwidth]{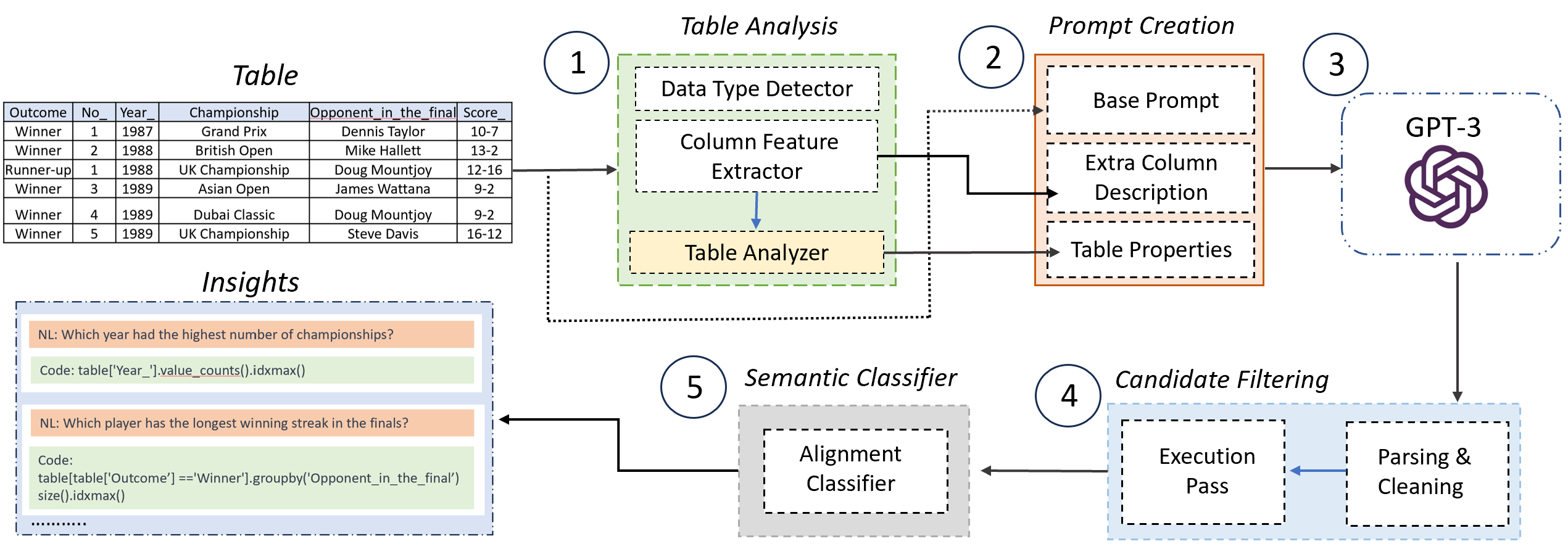}

\caption{Overview of approach. \circled{1} The table is analysed to gather relevant information to \circled{2} build a prompt that helps the model \circled{3} to provide interesting insights. \circled{4} The output of the model is cleaned and invalid suggestions are filtered. \circled{5} The suggestions are then screened by the Semantic Classifier to present the actionable insights.}
\label{fig:arch}
\end{figure*}

Given a table $T$, the goal is to find question and program pairs $(q_i, p_i)$.
We make sure that the generated pairs $(q_i, p_i)$ are aligned and $p_i(T)$ correctly answers $q_i$. Additionally, we need to ensure that we maintain a balance between diversity (unique $(q_i, p_i)$ pairs) and complexity ($q_i$ is non-trivial and provides an interesting insight). We make no assumption of the type of the data present inside each cell of the given table and consider it to be string.

\subsection{Architecture}

Figure~\ref{fig:arch} shows an overview of our approach.
\circled{1} We analyse the given table to \circled{2} build a prompt that contains relevant information about the table, and we \circled{3} instruct the model to generate questions and code pairs.
\circled{4} We then clean invalid and non-executable candidates, and finally \circled{5} remove those candidates where the generated code does not align with the question.
These stages of our approach are detailed in the following sections.

\subsection{Table Analysis and Prompt Creation}

The prompt contains three main relevant pieces of information, which together allow the model to perform the right wrangling operations when providing the code for insightful questions.

First, a small subset of rows allows the model to understand the syntax of each column.
For example, in Figure~\ref{tab:motiv_scenario}, an example of the score ``10--7'' is required to understand that splitting on ``--'' and integer conversion is required to compute averages.

Second, properties about columns provide the model with a complete picture about the scope of the data.
These properties are the appropriate data-type information (integer, float, string, and date-time) and number of missing elements of each column, and cardinality, extrema, peak-frequency, column position and entropy for each column. 

Third, we leverage these features to predict columns that are suitable candidates for grouping and aggregation.
A similar approach was used by Auto-Suggest~\cite{yan2020auto} for generating such operations, and we use their dataset for training the model.
Instead of complex post-processing, we only use the predictor to provide a weak signal to the LLM, which then leverages its semantic understanding to refine these candidates.
Our predictor is a decision tree, which achieves 88\% accuracy on an 80/20 split.

\subsection{Filtering and Semantic Alignment}

We instruct the model to generate question and code pairs and filter pairs for which $p_i$ cannot be executed on the input table.
Inspecting the resulting code pairs, we discovered that the question and code do not always align well together.

For example, consider the generated insight for the snooker data from Figure~\ref{tab:motiv_scenario}: \emph{Which player has the longest winning streak in the finals?}
\begin{minted}[fontsize=\small,breaklines]{Python}
> table[table['Outcome'] == 'Winner']
   .groupby('Opponent in final')
   .size().idxmax()
"Steve Davis"
\end{minted}
While this is an interesting question, the corresponding code does not take consecutive years into account. 
Similarly, consider the generated insight for another table with columns [year, starts, wins, poles, position, teams]: \emph{What is the average number of starts by drivers in the top 5 positions?}
\begin{minted}[fontsize=\small,breaklines]{Python}
>table[table['Position']
   .str.contains('5th')]['Starts']
   .mean()
"23"
\end{minted}
The code only takes the $5^{th}$ position in account, rather than the top 5, and this question and code pair is not semantically aligned.

We propose to predict the semantic alignment of question-code pairs $(q, p)$ by leveraging pre-trained embeddings of text and code \cite{embeddings}.
Instead of directly using the embeddings, we train a small, fully connected network to predict semantic alignment from the concatenated embeddings of $q$ and $p$.
Binary cross entropy between the predicted and the ground truth label is used as loss function.

We use a dataset of natural language descriptions and pandas expressions \cite{Jigsaw22} to create a dataset of aligned and misaligned pairs.
By using GPT-3 to generate code from these descriptions and using the execution result, we obtain 2209 aligned and 422 misaligned pairs.
Next, we swap the code for two pairs $(q_1, p_1)$ and $(q_2, p_2)$ to generate additional misaligned pairs $(q_1, p_2)$ and $(q_2, p_1)$.
In total, this resulted in 1552 misaligned pairs.
The semantic alignment classifier achieves 95.6\% F1 score on 80/20 split.

\section{User Study}
Effective insights should be perceived as useful and relevant to data analysis tasks. Prior to evaluating the semantic alignment of generated question and code pairs, we ensure the insights generated by our approach, enable productive exploration of datasets and are sufficiently complex. To measure the quality of generated questions, we performed a user study where we asked annotators to rate insights and provide feedback.

\subsection{Methodology}

\subsubsection{Participants} We recruited 5 participants from within our organization, who have a background in computer science, and varying levels of experience with data analysis tasks. 

\subsubsection{Task} Participants were asked to examine a sample of 12 tables randomly selected from our dataset, and rate 6--7 insights generated for each table. To ensure that participants spent sufficient time on each table, we first asked them to describe five exploratory analysis questions they might have about the table. This enabled participants to get acquainted with the data, and identify meaningful metrics and columns of interest from the data. Next, for each of the 76 insights presented, participants answered each of the following questions on a 7-point (1--7) Likert-scale:

\begin{enumerate} [leftmargin=4em, label=(M\arabic*)]
    \item This insight is relevant and enhances my understanding of the table.
    \item This insight saves time and improves my productivity.
    \item I would not have been able to independently arrive at a similar query as suggested by this insight.
\end{enumerate}

Ratings provided for these questions help us assess the relevance and usefulness of generated questions. Once the participants have provided ratings for all 76 insights, we ask them for their overall feedback and additional comments.

\subsubsection{Analysis}

\label{subsubsection:analysis_userstudy}
To analyze the ratings and assess which insights are considered meaningful by the participants, we use three metrics, one corresponding to each Likert-scale question: Relevance (M1), Productivity (M2), and Ingenuity (M3). We then rank all insights using a combined metric:
\[
Relevance + Productivity + Ingenuity
\]
We qualitatively analyze the ranked insights to identify desirable properties that contribute to increased usefulness of insights, and properties that lead to lower ratings.  

\subsection{Findings}

\definecolor{DarkGreen}{HTML}{013220}
\definecolor{Green}{HTML}{0f7f12}
\definecolor{LightGreen}{HTML}{93ec94}
\definecolor{LightPink}{HTML}{ffb6c1}
\definecolor{Pink}{HTML}{fc0d1b}
\definecolor{Red}{HTML}{8B0000}

\newcommand{\likert}[6]{
\begin{minipage}[l]{\columnwidth}
  \begin{tikzpicture}[xscale=0.0125, yscale=0.3]
      
    \node at (-100,0) {};
    \node at (100,0) {};

    \filldraw[color=Red] (#1, 0.0) rectangle (#2, 1.0);
    \filldraw[color=Pink] (#2, 0.0) rectangle (#3, 1.0);
    \filldraw[color=LightPink] (#3, 0.0) rectangle (0, 1.0);
    \filldraw[color=LightGreen] (0, 0.0) rectangle (#4, 1.0);
    \filldraw[color=Green] (#4, 0.0) rectangle (#5, 1.0);
    \filldraw[color=DarkGreen] (#5, 0.0) rectangle (#6, 1.0);

    \draw (0,0) -- (0, 1);
    
  \end{tikzpicture}
\end{minipage}    
}

\begin{table*}[!h]
\centering
\caption{Study Survey Responses\label{table:usefulness}}
\begin{threeparttable}
\begin{tabularx}{\textwidth}{lrrrrrrrrX}
\toprule
& & \multicolumn{7}{c}{\textbf{Likert Response Counts$^1$}}\\
\cmidrule{3-9}
 & 
  \textbf{\% Agree} & 
  \textbf{SD} & 
  \textbf{D} &
  \textbf{SwD} &
  \textbf{N} &
  \textbf{SwA} &
  \textbf{A} & 
  \textbf{SA} & 
  \multicolumn{1}{c}{\textbf{Distribution$^2$}}
\\ \midrule
& & & & & & & & &
\begin{minipage}[l]{\textwidth}
  \begin{tikzpicture}[xscale=0.0125, yscale=0.3]    
    \node at (-100,0) {};
    \node at (100,0) {};
    \draw (-100,0) -- (100,0);
    \draw (-100,-0.25) -- (-100, 0.25);
    \draw (100,-0.25) -- (100, 0.25);    
    \draw (-50,-0.25) -- (-50, 0.25) node[above] {\tiny 50\%};
    \draw (50,-0.25) -- (50, 0.25) node[above] {\tiny 50\%};
    \draw (0,-0.25) -- (0, 0.25) node[above] {\tiny 0\%};    
  \end{tikzpicture}
\end{minipage}\\
The insight is relevant \& enhances my understanding. & 79.21\% & 12 & 30 & 17 & 20 & 67 & 100 & 134 & \likert{-15.5}{-12.4}{-4.5}{18}{44}{79} \\

The insight saves time \& improves my productivity. & 76.84\% & 14 & 35 & 15 & 24 & 62 & 105 & 125 & \likert{-17}{-13}{-4}{16.3}{44}{76.8} \\

I would not have independently arrived at a similar query. & 35.52\% & 77 & 64 & 59 & 45 & 48 & 55 & 32 & \likert{-52.6}{-32.4}{-15.5}{12.6}{27.1}{35.5} \\

\bottomrule
\end{tabularx}

\begin{tablenotes}
\item[1] Strongly Disagree (SD), Disagree (D), Somewhat Disagree (SwD), Neutral (N), Somewhat Agree (SwA), Agree (A), Strongly Agree (SA).  \\
\item[2] Net stacked distribution removes the Neutral option and shows the skew between positive and negative responses. \\
\tikz \filldraw[color=Red] (0,0) rectangle (5pt,5pt); Strongly Disagree, \tikz \filldraw[color=Pink] (0,0) rectangle (5pt,5pt); Disagree, \tikz \filldraw[color=LightPink] (0,0) rectangle (5pt,5pt); Somewhat Disagree, \tikz \filldraw[color=LightGreen] (0,0) rectangle (5pt,5pt); Somewhat Agree, \tikz \filldraw[color=Green] (0,0) rectangle (5pt,5pt); Agree, \tikz \filldraw[color=DarkGreen] (0,0) rectangle (5pt,5pt); Strongly Agree.
\end{tablenotes} 

\end{threeparttable}

\end{table*}

\subsubsection{Ratings for Questions/Insights} 76 insights generated over 12 tables by our method were inspected by five raters for perceived relevance, improvements in productivity, and ingenuity. We report the ratings obtained in Table~\ref{table:usefulness}. Participants positively rated the relevance of insights overall (79\% agreement, median 6), and find them to be time-saving in analyzing data (77\% agreement, median 6). Lastly, participants find the presented insights to be a mix of trivial and complex generations, as suggested by the low agreement on Ingenuity of insights (35.5\% agreement, median 3). 

\subsubsection{Qualitative findings} We qualitatively inspected the ranked insights (as discussed in \ref{subsubsection:analysis_userstudy}) to find desirable properties that make generated questions useful. We find three overarching categories of insights, and present representative examples for each stratum: 

The \textit{\textbf{top 20 percentile}} insights provide highly semantically relevant statistics about the table. These insights have high M1, M2, and M3 ratings. They present non-trivial code generations that use complex operations in moderately long chains (such as groupby, apply, and diff). These insights highlight the most critical data points and exploratory findings while having high entropy (in other words, being interesting to the raters).
Following are examples from this category of insights:

\noindent\textbf{How many locations have a capacity in use greater than 100\%} (Rank 7; Score 13.67)
\begin{minted}[fontsize=\small]{Python}
> len(table[table['Capacity_in_use']
        .str.rstrip('%
        > 100]
    )
\end{minted}
\begin{itemize}
    \item[$\rightarrow$] The LLM realizes that the data reports transport capacity and utilization across cities, and the insight brings to attention the cities with overburdened transportation facilities.  
\end{itemize}
\smallskip

\noindent\textbf{What is the range of municipality areas in each region? (max area - min area)} (Rank 8; Score 13.5)
\begin{minted}[fontsize=\small]{Python}
> table.groupby('Regional_County_Municipality')
               ['Area_km_2_']
       .apply(lambda x: x.max()-x.min()) 
\end{minted}
\begin{itemize}
    \item[$\rightarrow$] The insight groups the data by regions and uncovers the range of land area. Our model is able to pick the correct columns names, while presenting a complex and meaningful insight.
\end{itemize}
\smallskip

The next stratum of insights lying in \textbf{\textit{20--80 percentile}} have high M1 and M2 ratings, with low M3 ratings---these insights are relevant, and improve productivity, but are commonly thought of by the raters (indicated by low \emph{Ingenuity} ratings). These insights involve computation of meaningful but commonly reported statistics (such as min, max, avg, value counts, and top-n) after performing groupby operations, or conditional filtering of data. Some examples include:

\noindent\textbf{How many games have been released in all three regions?} (Rank 16; Score 13.00)
\begin{minted}[fontsize=\small]{Python}
> table.loc[(table['Europe'] != 'N/A N/A')
          & (table['North_America'] != 'N/A N/A')
          & (table['Japan'] != 'N/A N/A')]
       .shape[0])   
\end{minted} 
\begin{itemize}
    \item[$\rightarrow$] The insight presents a sufficiently complex code snippet, but is commonly thought of by users. It identifies that the table contains release dates in three countries and leverages missing values to count games available in all countries.
\end{itemize}
\smallskip

\noindent\textbf{What is the most common nationality among the directors who have won awards?} (Rank 41; Score 11.8)
\begin{minted}[fontsize=\small]{Python}
>>> table.groupby('Nationality_of_director')
        ['Award'].count().idxmax() 
\end{minted}
\begin{itemize}
    \item[$\rightarrow$] Though this question helps participants gain a highly meaningful insight about their data, it is commonly thought of. 
\end{itemize}
\smallskip

Upon analyzing the \textbf{\emph{bottom 20 percentile}} insights we find three sub-categories, exposing characteristics of insights that are undesirable for participants. These findings further motivate the need for a filtering mechanism:
\begin{itemize}
    \item Majority of the insights in this strata compute trivial statistical measures for the data (such as min, max, mean, sum, nunique). Participants provided low M1, M2, and M3 ratings for these questions. For instance:
    
    \noindent\textbf{How many unique NFL teams are present in the data?} (Rank 62; Score 10.16)
    \begin{minted}[fontsize=\small]{Python}
    > table['NFL_Team'].nunique()
    \end{minted}
    \smallskip
    
    \item Few insights contained highly instance-specific conditions or constraints to filter data, and are not perceived to be useful despite performing complex operations. Such insights also receive low ratings on M1, M2, and M3. For instance:
    
    \noindent\textbf{What is the highest position secured by Scissor Sisters in the chart?} (Rank 76; Score 8.00)
    \begin{minted}[fontsize=\small,breaklines]{Python}
    > table[table['Artist']
               =='Scissor Sisters']
           ['Highest_Position'].max() 
    \end{minted}
    \begin{itemize}
    \item[$\rightarrow$] Here, participants did not find much use in insight finding for `Scissor Sisters' in particular, and expected the insight to be more generic.
\end{itemize}
    \smallskip
    
    \item Lastly, two insights could not be drawn from the provided table, and their code generations used incorrect target column(s). For instance, one of these insights attempts to report information on population age, while the table does not contain any columns that indicate age:  
    \noindent\textbf{Which region has municipalities with the highest and lowest average population age?} (Rank 68; Score 10.33)
    \smallskip
\end{itemize}

\subsubsection{End-of-study survey} Ratings from the end-of-survey questions show that 4 \textit{of} 5 participants found it challenging to get acquainted with the data, and identify meaningful insights that can be derived --- \emph{``It is difficult to think of meaningful queries without context of the data. While the tables in most cases were self-explanatory, there were a few which I just didn't understand. I think generated insights were very helpful in this''} (R3). Responses from the open-field feedback suggest that insights generated using our method help participants by (1) enhancing their understanding of the data, (2) identifying columns of interest, and (3) suggesting non-trivial and semantically relevant insights. 

\section{Evaluation}

We conducted an empirical study to answer the following research questions:
\begin{itemize}[leftmargin=3em]
    \item[\bfseries RQ1] Can we leverage embeddings to select semantically aligned (question, code) pairs from model generations?
    \item[\bfseries RQ2] How do the generation task, number of insights and prompt influence the diversity of insights?
    
\end{itemize}

We used the Open-WikiTable corpus~\cite{kweon2023open} for our experiments.
The dataset contains 24,680 tables, and natural language questions. The dataset was specifically constructed for the purpose of evaluating systems that perform complex reasoning tasks over tables. %
To ensure our evaluation would be tractable, we randomly selected 430 tables from the 24,680 tables available.

\subsection{Semantic Alignment (RQ1)}

We evaluate whether embeddings can be used to detect semantic alignment, and perform an ablation on different strategies of combining the embeddings for question and code.

\subsubsection{Experimental Setup}

We compare our semantic alignment classifier against data annotated by humans.

To create our annotated dataset, we first generated 10,175 insights across 430 tables---about 25 insights per table. For some tables, fewer insights were produced when token limits were exceeded. We were able to execute 8,954 out of 10,175 (88\%) of the generated insights.
From the executable insights, a sample of 309 (question, code) pairs---stratified on code length---was annotated by expert data scientists as semantically aligned or not.

We use the OpenAI \texttt{\small gpt-3.5-turbo}, \texttt{\small gpt-3.5-turbo-16k}, and \texttt{\small gpt-4} models as baselines for comparison.
Table \ref{tab:annotations} shows individual annotation statistics and an ensemble that only considers alignment if all other annotators (human and LLM) agree.

\begin{table}[htb]
    \centering
    \caption{Annotation dataset with 309 data points amongst different annotators.}
    \label{tab:annotations}
    \begin{tabular}{lrr}
        \toprule
        \textbf{Annotator} & \textbf{Positive} & \textbf{Negative}  \\\midrule
        Human & 251 & 58 \\ 
        GPT-4 & 222 & 87  \\ 
        GPT-3.5 & 230 & 79\\  
        GPT-3.5 16 K &228 & 81 \\ 
         Ensemble & 173 & 136 \\ 
        \bottomrule
    \end{tabular}
\end{table}

Besides concatenating individual embeddings, we also consider two settings to test performance: (1) classification using a single embedding for question and code, and (2) applying two fully connected networks to the individual embeddings and measuring the cosine similarity between the projected embeddings. We show the performance of each technique on the annotated data.

\begin{figure}[thb]
        \centering
            \includegraphics[width=\columnwidth]{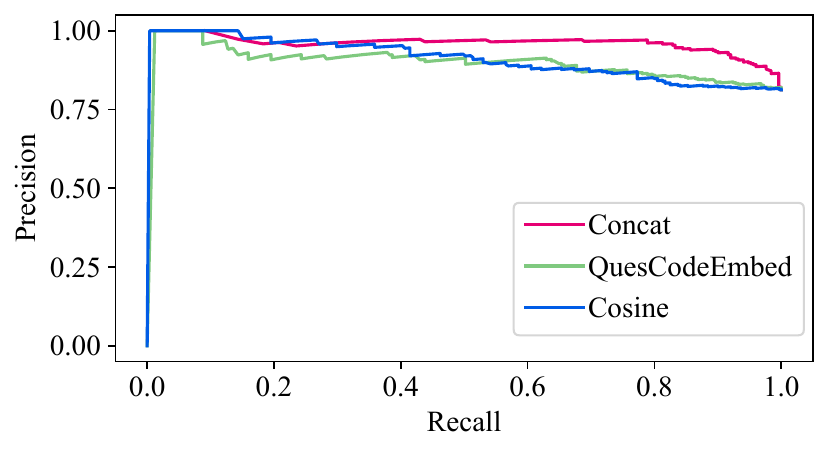}
            
    \captionsetup{justification=justified}
        \caption{
            PR curve showing performance of three different embedding variations, on annotated dataset.
        }
        \label{fig:prcurve_augment}
\end{figure}

\subsubsection{Results}

Figure~\ref{fig:prcurve_augment} shows precision-recall (PR) curves for different variations of the embedding models on human labelled data.
It highlights that concatenating embeddings (Concat) retains the highest precision for more recall.
At a recall of 80\%, this variation is 5.74\% and 6.8\% more precise than the embedding question and code together and the cosine similarity model, respectively.
We hypothesize that full attention between both embeddings allows the classifier to capture the most interactions.

Table~\ref{tab:Agreement_between_annotators} shows agreement between different models (our classifier and LLM baselines) and the human annotations, showing that our classifier (90\%) performs on par with GPT-4 (89.2\%).
Our embedding-based classifier is 728 times more cost effective\footnote{Currently, the average embedding computation cost per question and code pair is $0.007$ cents in comparison to $5.1$ cents by \texttt{\small gpt-4}. (7$^{\mathrm{th}}$ December 2023)} than GPT-4. 

\begin{table}[thb]
    \centering
    \caption{Agreement between models and human annotator. $^{**}$ best setting (Concat).}
    \label{tab:Agreement_between_annotators}
    \begin{tabular}{lrrrrrr}
        \toprule
        \textbf{Classifier} & \textbf{TP} & \textbf{TN} & \textbf{FP}& \textbf{FN} & \textbf{Acc.} & \textbf{F1}\\
        \midrule
        GPT-4 & 211 & 47 & 11 & 40 & 83.5\%  & 89.2\%\\ 
        GPT-3.5 & 200 & 27 &30 & 51 & 73.7\% & 83.1\%\\  
        GPT-3.5 16K &198 & 27&31 & 53 & 72.8\% & 82.5\%\\ 
        \textbf{Embed}$^{**}$ & 217 & 46 &12 & 34 & \textbf{85.1}\% &\textbf{ 90.0}\%\\ 
        \bottomrule
    \end{tabular}
\end{table}

\begin{figure}[tb]
\centering
\includegraphics[width=\columnwidth]{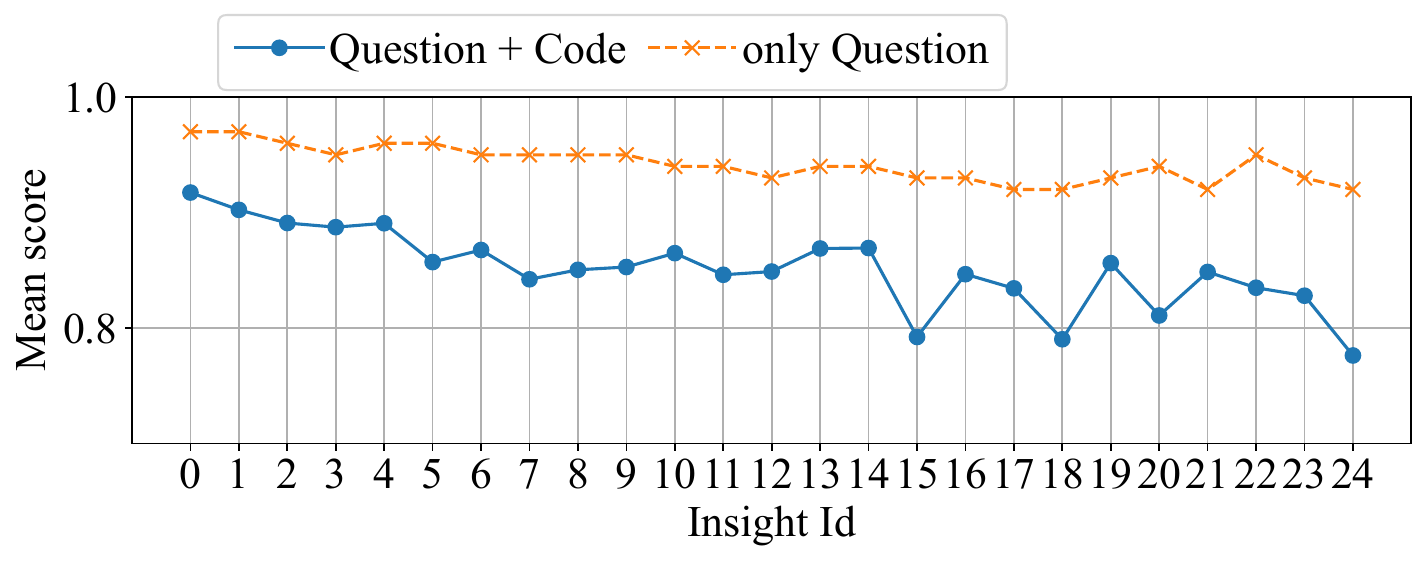} \caption{Distribution of semantic alignment between generated question and code pairs across different insights for tables.}
\label{fig:agreement}
\end{figure}

Figure~\ref{fig:agreement} shows how the alignment (logits from the classification model) slightly degrades as more insights are generated. 
We observed that the complexity of the insights increases as they were generated, which increases the potential for semantic alignment errors.
For example, an initial insight generated for the table in Figure~\ref{tab:motiv_scenario} is \emph{Which year had the highest number of championships?}
\begin{minted}[fontsize=\small,breaklines]{Python}
> table['Year'].value_counts().idxmax()
\end{minted}
As we generate more insights, the complexity increases.
Two examples are \emph{Which position has the highest number of goals in the league?}
\begin{minted}[fontsize=\small,breaklines]{Python}
> table.groupby(['Position_'])
      .agg({'League_Goals':sum})
      .sort_values(by=['League_Goals'],
                   ascending=False)
      .iloc[0].name
\end{minted}
and \emph{What is the most common party of Presidents who served multiple terms?}
\begin{minted}[fontsize=\small,breaklines]{Python}
> table[table['Name_'].isin(
    table['Name_'].value_counts()[
      table['Name_'].value_counts()>1
    ].index)
  ].groupby('Party')['Name_']
   .count()
   .sort_values(ascending=False)
   .iloc[0]
\end{minted}
Complex questions typically require grouping, sorting and filtering.

By manually inspecting cases where our classifier disagreed with human labels, we encountered situations where the distinction between aligned and misaligned was blurry.
This suggests that additional context might be necessary to identify alignment.
For example: \emph{Which universities had the highest number of students sign up in a given year?}
\begin{minted}[fontsize=\small,breaklines]{Python}
> table.loc[table.groupby('Year')['#SignedUp']
                 .idxmax()]
           [['Year', 'University', '#SignedUp']]
\end{minted}
The classifier identifies the above case as aligned, while humans disagree due to the presence of additional information (year and number of sign-ups). 

\subsubsection{Execution Prediction}

Executing code generated by the language model is not desired in production settings.
We therefore investigate whether the semantic alignment classifier can detect if code will execute or not.

\begin{figure}[tb]
\centering
\includegraphics[width=\columnwidth]{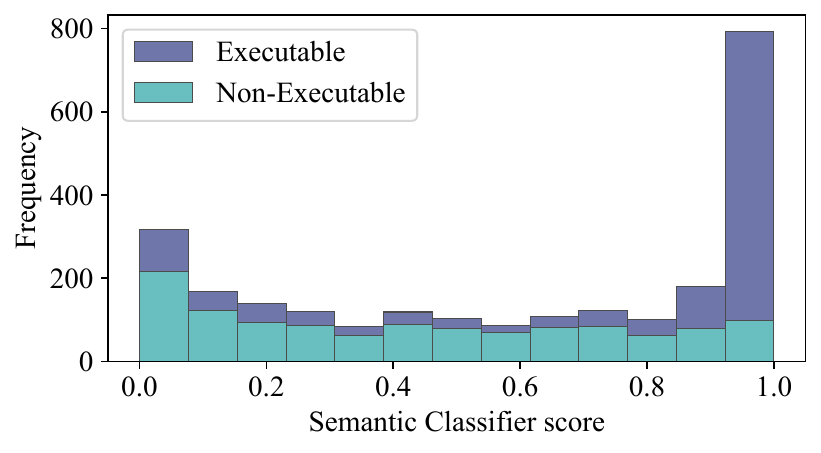} \caption{Distribution of semantic alignment scores among the executable and non-executable question code pairs stacked after one after another. }
\label{fig:SA_score_non_exec}
\end{figure}

Figure~\ref{fig:SA_score_non_exec} shows the logits distribution of the classifier over both executable and non-executable pairs.
Note that the semantic alignment is not known for executable pairs.
In general, non-executable pairs are given lower scores---scores close to zero occur the most often.
High alignment scores still occur for non-executable scores due to subtle syntax errors, which the embeddings do not capture.
Consider the following insight:
\emph{Who played the character with the maximum duration on the show?}
\begin{minted}[fontsize=\small,breaklines]{Python}
> table.iloc[table['Duration']
       .apply(lambda x: sum(
          int(i.split('x')[1]) - 
          int(i.split('x')[0]) + 1
          for i in x.split('-'))
       .idxmax()]['Actor']
\end{minted}
The code looks correct, but a parenthesis is unmatched, causing an alignment score of 0.9.

\subsection{Diversity (RQ2)}

We evaluate how different generation tasks, few-shot prompting and the number of insights influences the diversity of generations.

\subsubsection{Experimental Setup}

We measure the diversity of two (question, code) pairs by masking constants in the code and computing the edit distance on $8,954$ executable insights generated across 430 tables. For example,
\begin{center}
\begin{minted}[fontsize=\small]{Python}
table[table['Outcome']=='Winner']
    .groupby('Opponent in final')
\end{minted}
\end{center}
becomes \texttt{\small table[table[]==].groupby()}.
Diversity of multiple insights is computed as the average diversity across all pairs of code.

Besides generating questions and code together, we consider three variations: (1 and 2) generating only questions or code and performing the corresponding translation task, and (3) generating code and questions (in reverse order).
We use the \texttt{\small gpt-3} model again to generate the code for the given question and vice versa.
Only executable code generations are retained.

Apart from the zero-shot setup, we also try a one-shot (due to token limitations) setup with 3 static samples of question and code pairs.
The goal is twofold: showing the model what interesting questions look like, and ensuring that it follows the desired output format.

\subsubsection{Results}

\begin{figure}[bt]
    \captionsetup{justification=centering}
        \centering
        \begin{subfigure}{.49\columnwidth}
            \centering
            \includegraphics[width=\columnwidth]{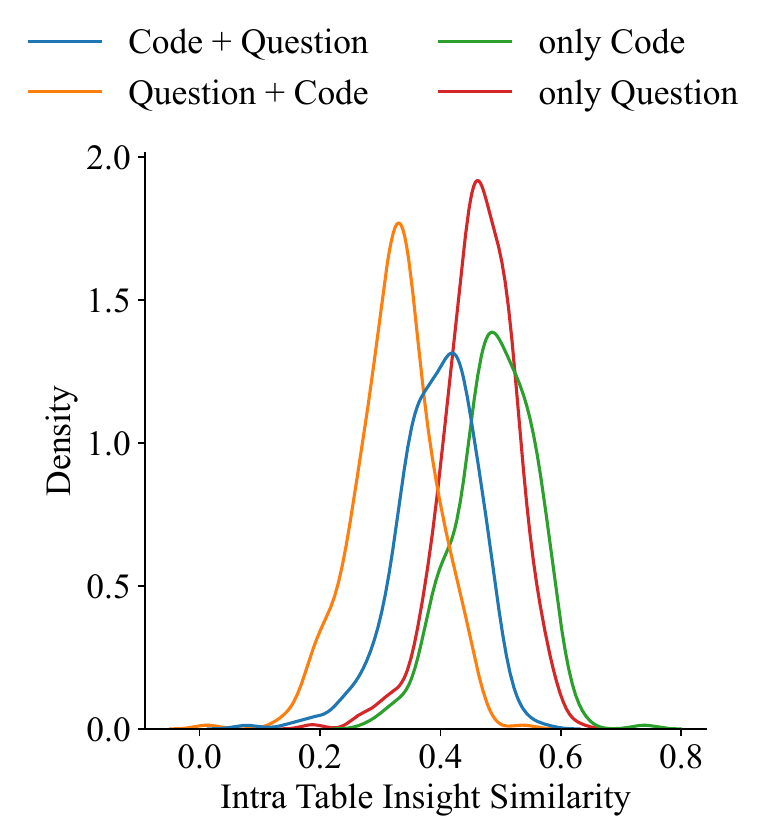}
            \caption{ Generation Style}
            \label{fig:diversity_accross_different_prompt_setup}
        \end{subfigure}
        \begin{subfigure}{.47\columnwidth}
            \includegraphics[width=\columnwidth]{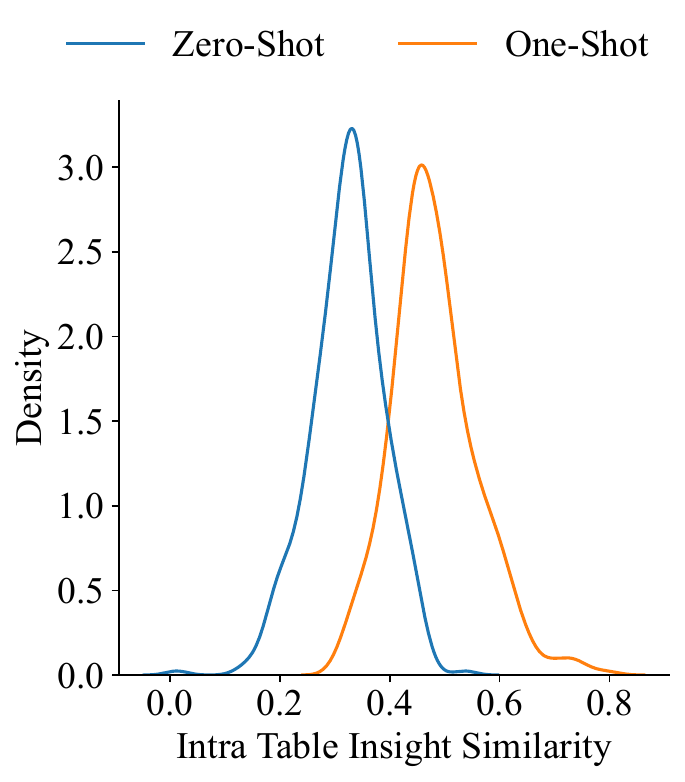}
            \caption{Prompting Technique}
            \label{fig:diversity_fewshot}
        \end{subfigure}
    \captionsetup{justification=justified}
        \caption{
            Variation in diversity across different (a) Generation styles (b) Prompting techniques. 
        }
        \label{fig:Diversity_across}
\end{figure}

Figure~\ref{fig:diversity_accross_different_prompt_setup} shows that generating questions and code together, in that particular order, yields more diverse questions than any other setup.
Generating a single modality (questions or code only) yields the least diversity.
We hypothesize that having code or questions closer together biases the model towards generating similar output.
Generating code only results in the least diverse questions.
When generating code and questions (in that order) the model likely pays more attention to the code, and again becomes more repetitive.

Figure~\ref{fig:diversity_fewshot} shows that providing an example of insights reduces the diversity.
We noticed that the model tries to mimic the given example for different columns, which harms the overall diversity.
For example, when providing \emph{What is the average salary amongst all participants?} as a seed insight, the model generates questions like \emph{What is the average number of matches played by each team?} and
\emph{What is the average difference between ``Points'' and ``Against'' for all the teams?}
In total, 9 out of 25 questions are about averages.

\begin{figure}[tb]
\centering
\includegraphics[width=\columnwidth]{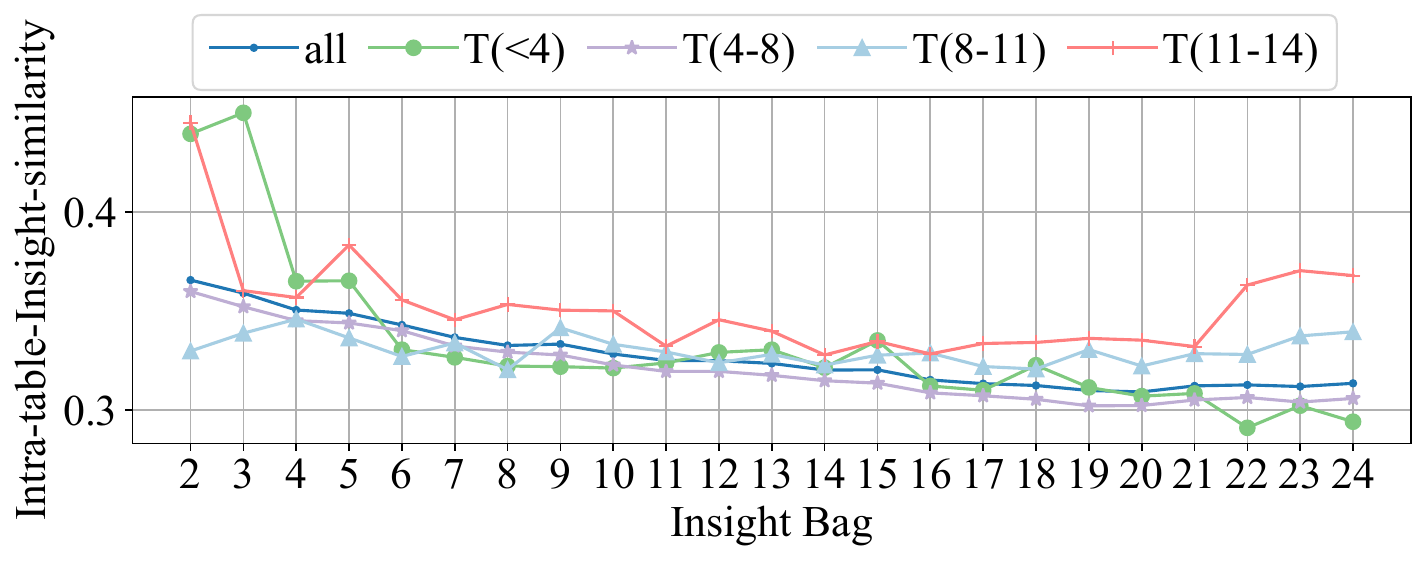} \caption{Evolution of diversity as more insights are generated for all tables and divided by number of columns. T(4-8) means tables with 4 to 8 columns. For tables with more columns, the model becomes repetitive.}
\label{fig:diversity_over_insights_col_number_var}
\end{figure}

\begin{table}[ht]
    \centering
    \caption{Token and time costs associated with different generation styles.}
    \label{tab:cost}
    \begin{tabular}{lcrrr}
        \toprule
        \textbf{Generation Style} & 
        \textbf{Tokens} & \textbf{Inference (in sec)} \\ \midrule
        Only Question &  
        56,382 
        & 121.52
        \\ 
        Only Code & 
        12,833
        & 62.3
        \\ 
        Code + Question & 
        1,512 &19.89  \\ 
        Question + Code & 
        1,515 & 20.9  \\  
        \bottomrule
    \end{tabular}
\end{table}

Figure~\ref{fig:agreement} shows that generating questions only and then translating to code yields higher alignment (93\%).
Two drawbacks of this approach are diversity and cost.
Table~\ref{tab:cost} shows the average number of tokens and associated inference time for different approaches.
Generating questions and code together is 37 times cheaper and six times faster than generation questions and translating them.

\begin{figure}[tb]
\centering
\includegraphics[width=\columnwidth]{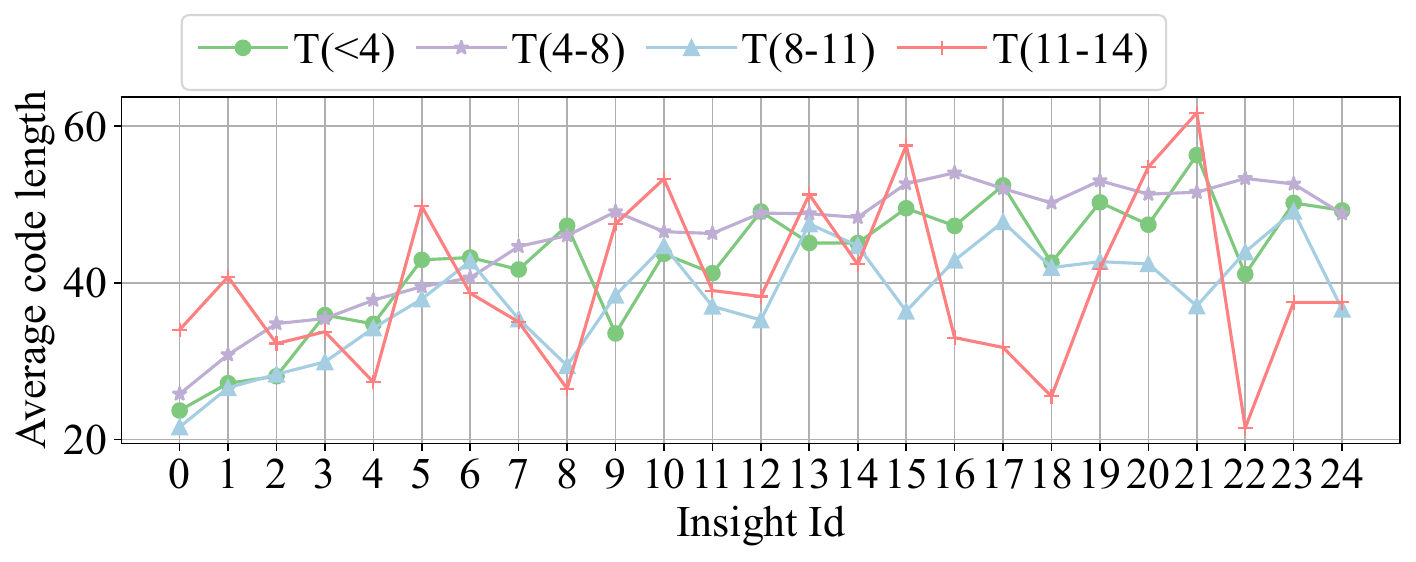}
\caption{Evolution of code length as more insights are generated. }
\label{fig:code-length-variation}
\end{figure}

Figure~\ref{fig:diversity_over_insights_col_number_var} shows that initial generations are diverse, but we see diminishing returns form generating more insights.
After around 20 insights, the questions becomes more repetitive.
Surprisingly, this effect is stronger for tables with more columns.
We find that the model repeats the the exact same insight for multiple columns.
To verify this effect, Figure~\ref{fig:code-length-variation} shows how the length of the code (after removing column names) evolves for tables with different numbers of columns. 
For smaller number of columns, the model cannot just repeat itself, and it generates more complex questions.

\section{Limitations}

In this work, we have focused solely on English text generation, leaving the exploration of suggestions in other languages as a potential avenue for future research.
Furthermore, we observed a notable drop in performance due to the data-distribution shift when testing the model from human annotated to machine-generated data.
It is essential to build models that can account for such variations.
Additionally, we treat the generated suggestions as an unordered set and do not rank or filter them based on diversity or relevance, which could be a promising direction for future investigation.
For generation tasks, we have only considered models from the \texttt{\small gpt-3} and \texttt{\small gpt-3.5} class, as these are state-of-the-art in generation capabilities.
However, the results may differ with smaller models, as they may lack in-context learning abilities.

\section{Conclusion}

In this paper, we explore leveraging LLMs to aid automated inisght generation, by using them to generate aligned  question and code pairs for tabular data.
Obtaining diverse and interesting questions, along with the code to generate them---even when transformations are required---is not possible with traditional methods, which rely on statistical features or historical operations from other users.
In our study, performed on data from Open-WikiTable, we showed that insights thus generated are meaningful and interesting to users. Further we showed that generating questions and code together yields diverse questions, and that an embedding-based semantic alignment classifier performs on par with GPT-4 for filtering cases where question and code are misaligned at a fraction of the cost.


\end{document}